 \documentclass[final,3p,times,twocolumn]{elsarticle}

\usepackage{amssymb}
\usepackage{bm}

\journal{Physica C Special Issue on Stripes and Electronic Liquid Crystals in Strongly Correlated Systems}

\begin{document}

\begin{frontmatter}



\title{Thermodynamic evidence for broken fourfold rotational symmetry in the hidden-order phase of URu$_2$Si$_2$}


\author{T. Shibauchi}
\ead{shibauchi@scphys.kyoto-u.ac.jp}
\author{Y. Matsuda}
\ead{matsuda@scphys.kyoto-u.ac.jp}

\address{Department of Physics, Kyoto University, Sakyo-ku, Kyoto 606-8502, Japan}

\begin{abstract}
Despite more than a quarter century of research, the nature of the second-order phase transition in the heavy-fermion metal URu$_2$Si$_2$ remains enigmatic. The key question is which symmetry is being broken below this ``hidden order'' transition. We review the recent progress on this issue, particularly focusing on the thermodynamic evidence from very sensitive micro-cantilever magnetic torque measurements that the fourfold rotational symmetry of the underlying tetragonal crystal is broken. The angle dependence of the torque under in-plane field rotation exhibits the twofold oscillation term, which sets in just below the transition temperature. This observation restricts the symmetry of the hidden order parameter to the $E^{+}$- or $E^{-}$-type, depending on whether the time reversal symmetry is preserved or not.
\end{abstract}


\end{frontmatter}


\section{Introduction}
\label{intro}

\begin{figure}[t]
\includegraphics[width=1.0\linewidth]{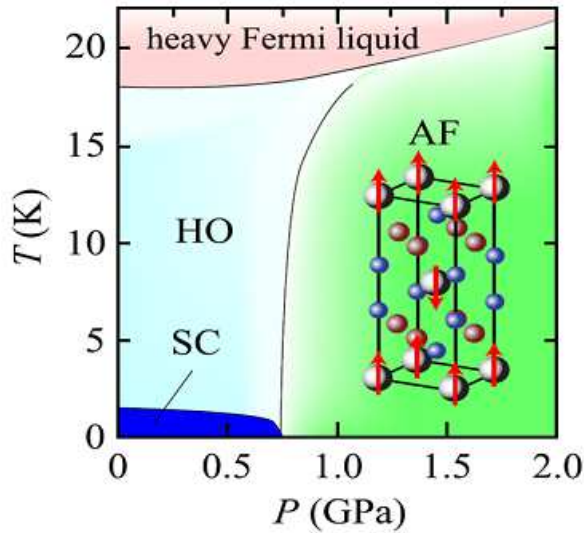}
\caption{Schematic temperature-pressure phase diagram of URu$_2$Si$_2$. At ambient pressure, the second-order phase transition from the paramagnetic, heavy Fermi liquid metallic state to the hidden order (HO) state occurs at $T_h=17.5$\,K. At lower temperature below $T_c=1.4$\,K superconducting (SC) state appears. Above the critical pressure ($\sim 0.75$\,GPa), the large-moment antiferromagnetic (AF) state sets in through the first-order phase transition. The inset illustrates the schematic crystal structure (large white spheres, middle red spheres, small blue spheres are U, Ru, and Si atoms, respectively) with the spin arrangements in the antiferromagnetic phase.
}
\label{TP}
\end{figure}

At $T_h=17.5$\,K, URu$_2$Si$_2$ undergoes a second-order phase transition accompanied by the large anomalies in thermodynamic and transport properties \cite{Pal85,Map86,Sch86,Ram92}. Several remarkable features have been reported in the lower-temperature hidden order phase \cite{Myd11}. URu$_2$Si$_2$ exhibits the body-centered tetragonal crystal structure [see inset of Fig.\,\ref{TP}], and to date no structural phase transition is observed at $T_h$. Tiny magnetic moment appears ($M_0 \sim 0.03\mu_B$) below $T_h$ \cite{Bro87}, but it is by far too small to explain the large entropy released during the transition and seems to have an extrinsic origin \cite{Mat01,Tak07,Ami07}.  The electronic excitation gap is formed at a large portion of the Fermi surface and most of the carriers ($\sim 90\%$) disappears \cite{Sch87,Beh05,Kas07}.  The gap is also formed in the incommensurate magnetic excitations below the hidden order transition \cite{Wie07}. With application of pressure, the large-moment antiferromagnetic state with the wave vector $\bm{Q}_c=(0,0,1)=(1,0,0)$ appears, which is separated from the hidden-order phase by the first-order phase transition [see Fig.\,\ref{TP}] \cite{Myd11,Ami07}. The quantum oscillation experiments \cite{Has10} show that the Fermi surface of the hidden-order phase is quite similar to that of the pressure-induced antiferromagnetic phase. These results suggest that the hidden order and antiferromagnetism are nearly degenerate. 

In general, a second-order phase transition causes a change in various type of symmetries, such as crystal, rotational, gauge and time reversal symmetries. An order parameter is introduced to describe the low-temperature ordered phase with reduced symmetries. Therefore the key to the nature of hidden order lies in understanding which symmetry is being broken. Several microscopic models, including multipole ordering \cite{San94,Kis05,Hau09,Cri09,Har10}, spin-density-wave formation \cite{Ike98,Min05,Elg09}, orbital currents \cite{Cha02} and helicity order \cite{Var06}, have been proposed.  However, in spite of the intensive experimental and theoretical studies, what is the genuine order parameter in the hidden order phase is still an open question.

\section{Searching for fourfold rotational symmetry breaking}
\label{rotational}

\subsection{Information from the superconducting state embedded in the hidden-order phase}

\begin{figure*}[t]
\begin{center}
\includegraphics[width=0.8\linewidth]{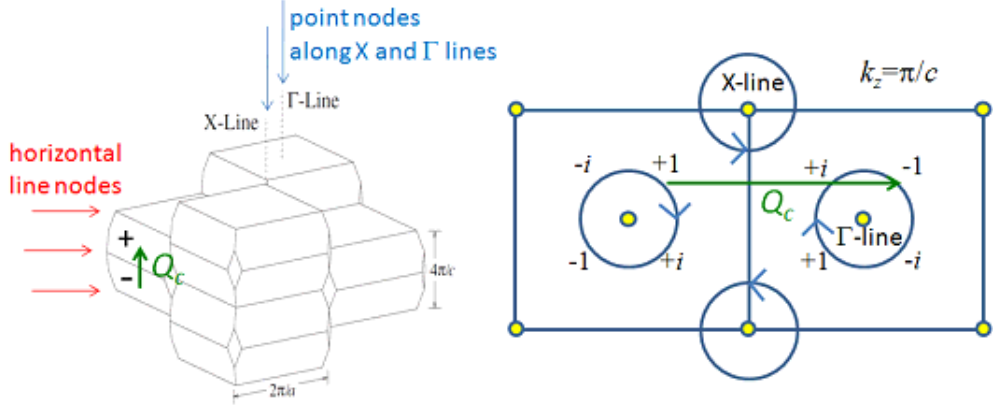}
\end{center}
\vspace{-5mm}
\caption{Schematic arrangements of the Brillouin zone for the body-centered tetragonal structure of URu$_2$Si$_2$ (left). The superconducting order parameter with chiral $d$-wave symmetry described by Eq.\,(\ref{chiral}) has horizontal nodes in the planes shown by the red arrows as well as point nodes along the X and $\Gamma$ lines indicated by the blue arrows. A cross sectional cut at $k_z=\pi/c$ including two neighboring Brillouin zones is also depicted (right) with the phase winding directions (circular arrows) for one of the degenerate chiral $d$-wave order parameters. The other order parameter has the opposite phase winding directions in each position. At the two points connected by the antiferromagnetic wave vector $\bm{Q}_c$ (green arrows), which is $(0,0,1)$ (left) or $(1,0,0)$ (right), the superconducting order parameter changes sign.
}
\label{BZ}
\end{figure*}

Inside the hidden-order phase, the unconventional superconductivity appears below $T_c=1.4$\,K. When the antiferromagnetic phase is induced by the pressure, the superconductivity disappears [see Fig.\,\ref{TP}]: i.e. the superconductivity can coexist with the hidden order, but not with the antiferromagnetic order. Therefore the information on the superconducting state is also important for understanding the hidden order. The thermal conductivity \cite{Kas07,Kas09} and specific heat \cite{Yan08} measurements using very clean single crystals, which become available recently \cite{Mat11}, consistently show that the field dependence of these quantities has strong anisotropy. From these results, both suggest the existence of point nodes along the $c$ axis in the superconducting order parameter. The symmetry analysis for the spin-singlet states leads to the chiral $d$-wave symmetry with the form \cite{Kas07}
\begin{equation}
\sin\frac{k_z}{2}c\left(\sin\frac{k_x+k_y}{2}a\pm\mathrm{i}\sin\frac{k_x-k_y}{2}a\right)
\label{chiral}
\end{equation}
that has the zero points along the $\Gamma$ and X lines in the Brillouin zone as well as the horizontal line nodes as shown in Fig.\,\ref{BZ}. This symmetry of superconducting order parameter $\Delta(\bm{k})$ has sign change at the points connecting with the antiferromagnetic vectors $\bm{Q}_c=(0,0,1)$ and $(1,0,0)$, which are equivalent for the body-centered tetragonal structure. This sign-changing order parameter is consistent with the superconductivity mediated by the antiferromagnetic spin fluctuations. The commensulate antiferromagnetic fluctuations lead to the enhanced dynamical susceptibility at $\bm{Q}_c$ and hence the pairing interaction becomes large and repulsive at $\bm{Q}_c$, leading to the sign change of the superconducting order parameter $\Delta(\bm{k})=-\Delta(\bm{k}+\bm{Q}_c)$ \cite{Hir11}. The disappearance of superconductivity in the pressure-induced antiferromagnetic phase implies that the $\bm{Q}_c$ spin fluctuations are suppressed by the antiferromagnetic ordering. Recent theoretical calculations also point to the chiral $d$-wave superconductivity produced by the antiferromagnetic fluctuations \cite{Kus11}.

The $\pm$ signs in Eq.\,(\ref{chiral}) stems from the degenerate nature of the two states under the tetragonal symmetry in the system. The two states are characterized by the phase winding directions of the order parameter in the $ab$ plane as shown in Fig.\,\ref{BZ}. Recent lower critical field $H_{c1}(T)$ measurements detect some unusual anomaly below the superconducting transition \cite{Oka10}. From this observation, a possibility of the phase transition inside the superconducting state due to the degeneracy of the chiral $d$-wave has been pointed out. Namely, if the hidden order state breaks the fourfold rotational symmetry, the superconducting states with two different chiralities (the states with $+$ and $-$ sign in the Eq.\,(\ref{chiral}) form) may have different transition temperatures. Although the origin of the $H_{c1}(T)$ anomaly should be scrutinized, this has triggered the search for the in-plane fourfold symmetry breaking in the hidden order phase.

\subsection{Magnetic torque as a sensitive probe for the rotational symmetry breaking}

\begin{figure*}[t]
\includegraphics[width=1.0\linewidth]{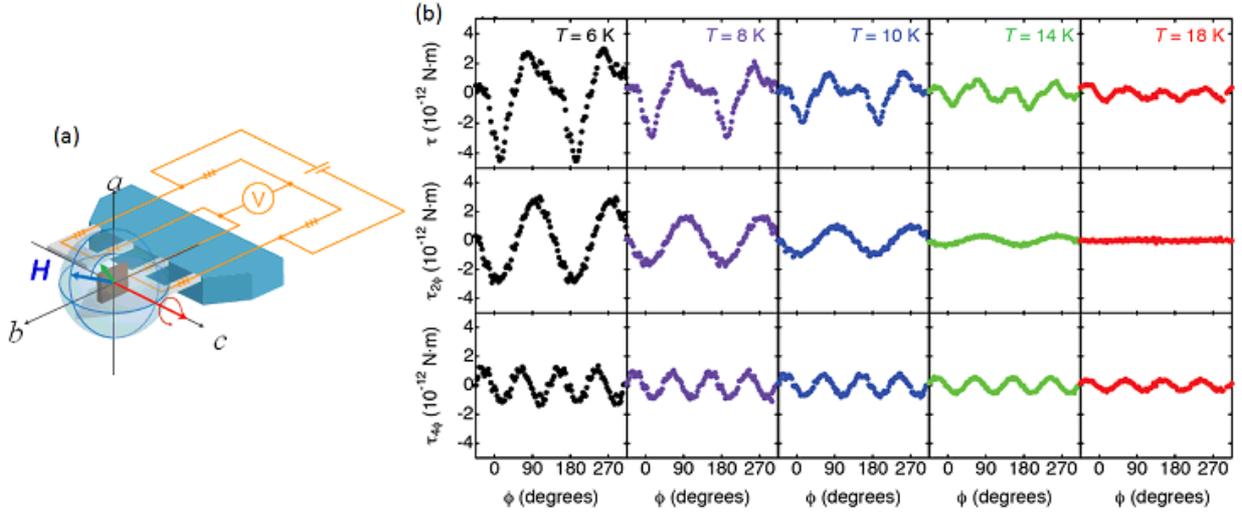}
\caption{(a) Schematic configuration of the magnetic torque measurements for $ab$-plane field rotation by using the micro-cantilever technique. The magnetic field $\bm{H}$ (blue arrow) induces the magnetization $\bm{M}$ (green arrow) in the URu$_2$Si$_2$ crystal. The torque along the $c$ axis (red arrow) can be detected by the change in the piezo resister, which is measured by the bridge configuration. (b) Upper panels show raw magnetic torque curves as a function of the azimuthal angle $\phi$ at several temperatures. All data are measured at $\left|\mu_0\bm{H}\right| = 4$\,T. Middle and lower panels show twofold $\tau_{2\phi}$ and fourfold $\tau_{4\phi}$ components of the torque curves which are obtained from the Fourier analysis.
}
\label{tau}
\end{figure*}

Among experimental probes, the magnetic torque $\bm{\tau} = \mu_0\bm{M}V \times \bm{H}$ is a particularly sensitive probe for detecting the magnetic anisotropy, where $V$ is a sample volume and $\bm{M}$ is the induced magnetization. The torque is also a thermodynamic quantity which is the angle derivative of free energy. The torque in tiny crystals can be measured by using the micro-tip cantilever \cite{Ros96,Ohm02}, which has a sensitivity orders of magnitude larger than the commercial SQUID magnetometer. This technique has been widely used for the de Haas-van Alphen quantum oscillation measurements at high magnetic fields or for determination of the anisotropy parameter in layered superconductors. We use this technique to measure in-plane anisotropy of magnetic susceptibility in the hidden-order phase \cite{Oka11}. The torque measurements in magnetic fields $\bm{H}=(H\cos\phi,H\sin\phi,0)$ rotating within the tetragonal $ab$ plane in URu$_2$Si$_2$ (where $\phi$ is the angle from the $[100]$ direction) provide a stringent test whether the hidden order parameter breaks the crystal fourfold symmetry. In such a geometry $\bm{\tau}(\phi,T,H)$ has a twofold oscillation with respect to $\phi$-rotation 
\begin{equation}
\tau_{2\phi} = \frac{1}{2}\mu_0H^2V\left[(\chi_{aa}^{}-\chi_{bb}^{})\sin2\phi - 2\chi_{ab}\cos2\phi\right],
\end{equation}
where the susceptibility tensor $\chi_{ij}$ is given by $M_i=\displaystyle\sum_j \chi_{ij} H_j$ [see Fig.\,\ref{tau}(a)].  

It should be stressed that in a system holding the tetragonal $C_4$ symmetry, $\tau_{2\phi}$ should be zero because $\chi_{aa}=\chi_{bb}$ and $\chi_{ab}=0$. Even when we consider non-linear (higher order) susceptibility, the $C_4$ symmetry allows only fourfold oscillation with respect to $\phi$-rotation, and cannot lead to the twofold component. Finite values of $\tau_{2\phi}$ may appear if a new electronic or magnetic state emerges that breaks the tetragonal symmetry. In such a case, rotational symmetry breaking is revealed by $\chi_{aa}\ne\chi_{bb}$ or $\chi_{ab}\ne0$ depending on the direction of the orthorhombic anisotropy. 

It has been reported by several groups \cite{Uem05,Ami07} that URu$_2$Si$_2$ single crystals may contain ferromagnetic (FM) impurities. Indeed, some of our crystals also show the sign of the FM impurities: In these crystals, the torque curve in rotating field $\bm{H}=(H\sin\theta,0,H\cos\theta)$ within the $ac$ plane shows the hysteresis due to the FM impurities when the angle $\theta$ from the $c$ axis is close to 90$^\circ$. We have carefully selected small pure crystals, in which no such hysteresis is observed within the experimental resolution of $\sim 10^{-14}$\,Nm ($\sim 0.01$\% of the amplitude $|A_{2\theta}|$ of the oscillation due to the difference between $\chi_{cc}$ and $\chi_{aa}$). 

\begin{figure*}[t]
\begin{center}
\includegraphics[width=0.8\linewidth]{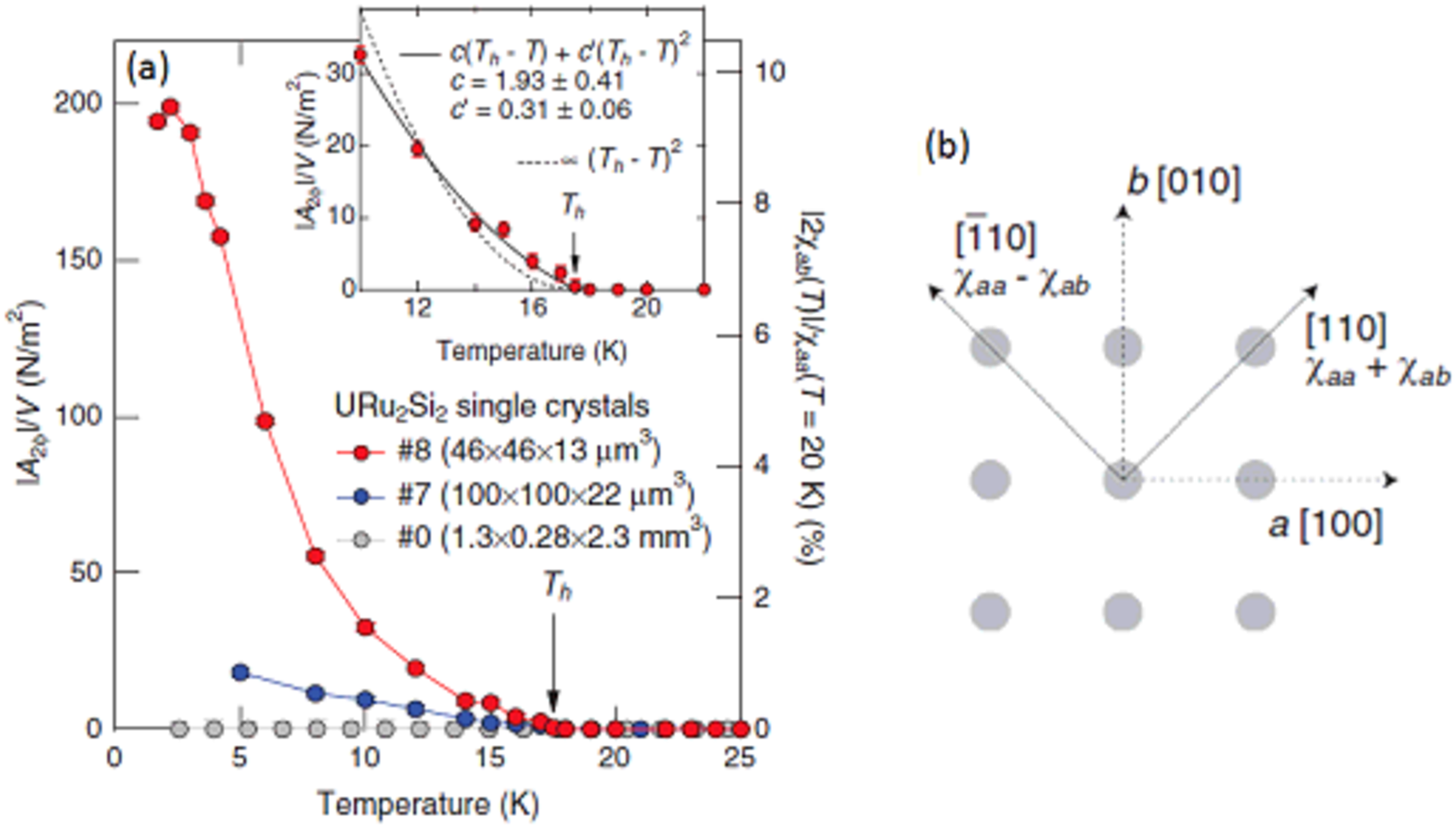}
\end{center}
\vspace{-5mm}
\caption{(a) Temperature dependence of the amplitude $|A_{2\phi}|$ of the twofold oscillations in $\tau(\phi)$ divided by the sample volume $V$ for two crystals \#8 (red circles) and \#7 (blue circles) with dimensions indicated in the figure. The corresponding susceptibility difference $|2\chi_{ab}|=|\chi[110]-\chi[\bar{1}10]|$ normalized by $\chi_{aa}$ is indicated in the right axis. The susceptibility difference measured by the SQUID magnetometer for the large crystal \#0 is also plotted (gray circles). (b) Schematic U atom arrangement (gray circles) in the $ab$ plane. The emergent nonzero $\chi_{ab}$ is indicative of non-equivalent susceptibilities along $[110]$ and $[\bar{1}10]$ directions (arrows).
}
\label{amplitude}
\end{figure*}

We also carefully aligned the magnetic field within the plane by using a system with two superconducting magnets generating $\bm{H}$ in two mutually orthogonal directions and a $^3$He cryostat equipped on a mechanical rotating stage with a minimum step of $1/500$\,deg \cite{Iza01}.  By computer controlling two magnets and the rotating stage, we were able to rotate $\bm{H}$ continuously within the $ab$ plane with a misalignment less than 0.02\,deg from the $ab$ plane. With this technique, we were able to observe no twofold oscillations in $\tau(\phi)$ measurements in the paramagnetic phase above $T_h$ up to 40\,K. 

\section{Thermodynamic evidence for fourfold symmetry breaking}
\label{results}

\subsection{Twofold oscillation in the in-plane magnetic torque}

Figure\,\ref{tau}(b) displays the evolution of the in-plane torque $\tau(\phi)$ at $|\mu_0\bm{H}|=4$\,T on entering the hidden order phase \cite{Oka11}.  At all temperatures, torque curves are perfectly reversible with respect to the field rotation directions.   Moreover, the torque curves remain unchanged for field cooling conditions with different field angles. $\tau(\phi)$ can be decomposed as $\tau=\tau_{2\phi}+\tau_{4\phi}+\tau_{6\phi}+\cdots$, where $\tau_{2n\phi}=A_{2n}\sin 2n(\phi-\phi_0)$ is a term with $2n$-fold symmetry with $n=1,2,\cdots$.   In the middle and lower panels, the twofold and fourfold components obtained from the Fourier analysis are displayed. 

What is remarkable is that the twofold oscillation is distinctly observed in the hidden order state, while it is absent in the paramagnetic phase at $T = 18$\,K slightly above $T_h$, as shown in the middle panels of Fig.\,\ref{tau}(b). Indeed, the torque curves shown by the upper panels apparently become asymmetric with respect to $90^\circ$ rotations below $T_h$.  We note that the fourfold oscillations $\tau_{4\phi}$ (and higher-order terms) observed for all temperatures arise primarily from the higher-order nonlinear susceptibilities \cite{Mor84}.  We therefore focus our attention on the twofold oscillation.  Figure\,\ref{amplitude}(a) depicts the temperature dependence of the amplitude of the twofold symmetry $|A_{2\phi}|$.  The amplitude onsets precisely at $T_h=17.5$\,K [see inset of Fig.\,\ref{amplitude}(a)]. This result, together with no hysteresis in the torque curves, again rules out a possibility that very tiny ferromagnetic impurity is an origin of twofold symmetry. Based on these results, we conclude that the amplitude of twofold oscillations is a manifestation of intrinsic in-plane anisotropy of the susceptibility. The $\tau_{2\phi}$ term has the form $\cos 2\phi$, which indicates the emergence of non-zero $\chi_{ab}$ below $T_h$. This means 
\begin{equation}
\chi[110]=\chi_{aa}+\chi_{ab} \ne \chi[\bar{1}10]=\chi_{aa}-\chi_{ab},  
\end{equation}
as schematically shown in Fig.\,\ref{amplitude}(b). 

The in-plane anisotropy that sets in precisely at $T_h$ indicates that the rotational symmetry is broken in the hidden-order phase. The temperature dependence of $|A_{2\phi}| \propto \chi_{ab}$ near $T_h$ is fitted with $\chi_{ab} \propto c(T_{h}-T)+c'(T_{h}-T)^{2}$ much better than with a purely quadratic fit [see inset of Fig.\,\ref{amplitude}(a)]. The leading $T$-linear term near $T_h$ is naturally expected from the standard Landau theory of second-order transition when $\chi_{ab}$ is described by the squared term of an order parameter $\eta \propto (T_{h}-T)^{1/2}$. Thus, the observed in-plane anisotropy $2\chi_{ab}/\chi_{aa}$ implies that the hidden-order parameter $\eta$ breaks the fourfold tetragonal symmetry. 

\subsection{Domain formation due to the degenerate nature of nematicity}

The breaking of the $C_4$ rotational symmetry in a tetragonal crystal structure is in general a hallmark of electronic nematic phases \cite{Kivelson98}. In such nematic phases that reduce $C_4$ to $C_2$ symmetry, there are two degenerate directions to which the orthorhombic anisotropy is characterized. In the present case, the $[110]$ (or $[\bar{1}\bar{1}0]$) and $[\bar{1}10]$ (or $[1\bar{1}0]$) directions are energetically equivalent (corresponding to the positive and negative $\chi_{ab}$), which naturally leads to the domain formation in the hidden order phase [see Fig.\,\ref{amplitude}(b)]. 

The domain formation results in the crystal size dependence of the twofold oscillation amplitude in the torque experiment. When the crystal size is much larger than the domain size, the volume difference of the two different domains can be negligibly small compared with the total volume. Then the contributions of the two domains to the twofold oscillations in the torque have out of phase, which leads to the cancellation of the oscillations. Indeed, the torque measurements in a bigger crystals show that the normalized signal becomes smaller, and in the mm-sized crystals the bulk susceptibility measurements cannot detect any difference between $\chi[110]$ and $\chi[\bar{1}10]$, as shown in Fig.\,\ref{amplitude}(a). 
Our results can be understood if the domain size is of the order of tens of microns, which may explain the difficulties in observing this effect by using the conventional techniques. The fact that the torque curves remain unchanged for field cooling conditions with different field angles implies that formation of such domains is predominantly determined by and strongly pinned by the underlying crystal conditions, such as internal stress or disorder. 

The magnitude of the susceptibility difference $\chi[110]-\chi[\bar{1}10]=2\chi_{ab}$ deduced from the twofold oscillations in the smallest crystal used in the torque measurements is as large as 10\% of the in-plane susceptibility [see Fig.\,\ref{amplitude}(a)]. We stress that even in the small crystals such a large signal cannot be accounted for by some unknown effect of the surface layer, whose volume is of the order of $10^{-4}$ of the total volume. It is therefore unambiguous that the signal comes from the bulk of URu$_2$Si$_2$. 

\subsection{Field dependence of the twofold oscillations}

\begin{figure*}[t]
\begin{center}
\includegraphics[width=0.78\linewidth]{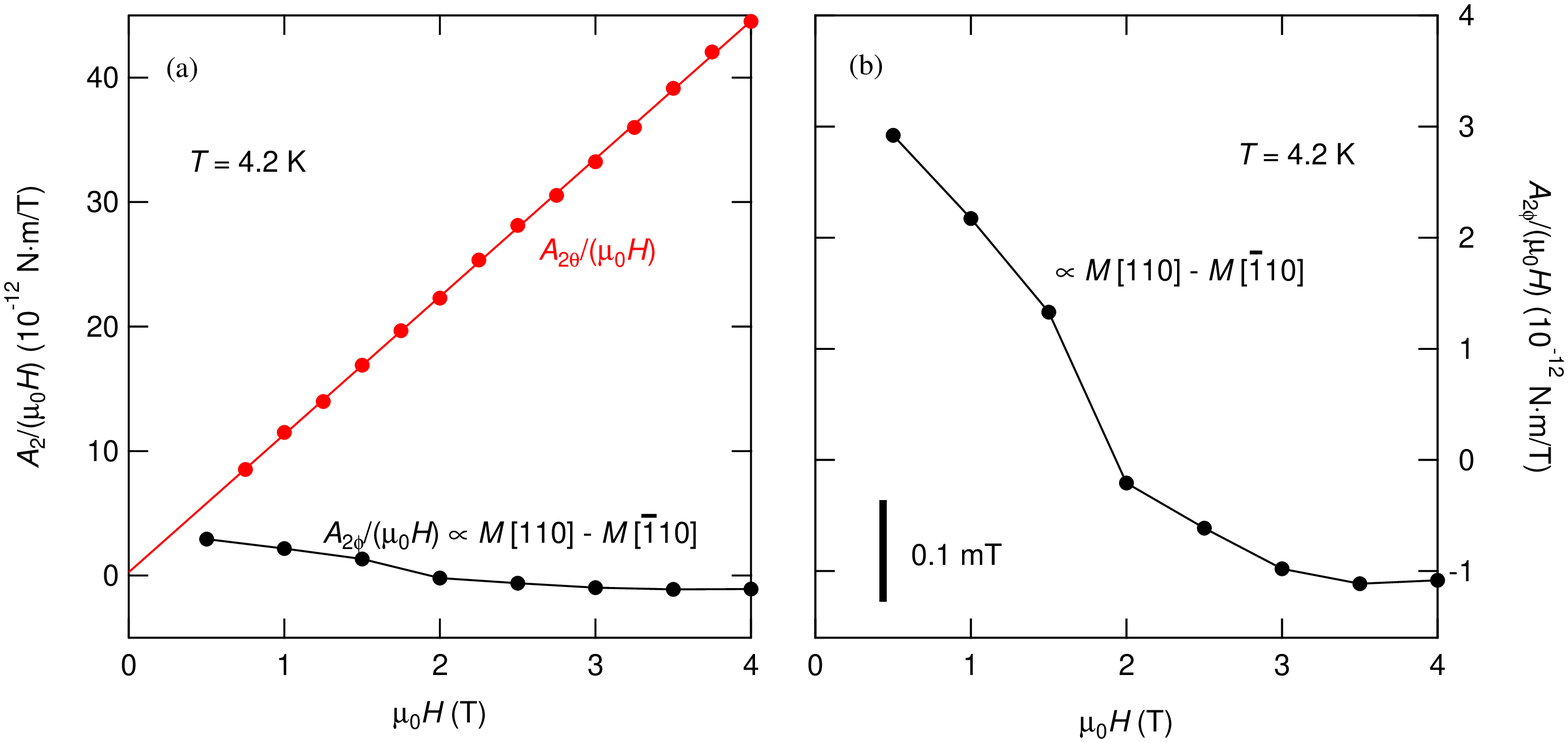}
\end{center}
\vspace{-5mm}
\caption{(a) Magnetic field dependence of the amplitudes of the twofold oscillations $A_{2\theta}/(\mu_{0}H)$ for the $ac$-plane rotation and $A_{2\phi}/(\mu_{0}H)$ for the $ab$-plane rotation. (b) Expanded view for $A_{2\phi}/(\mu_{0}H)$. 
}
\label{tau_H}
\end{figure*}

In Figs.\,\ref{tau_H}(a) and (b) we depict the amplitude of the twofold oscillation in $\tau(\phi)$ divided by the field $A_{2\phi}/(\mu_{0}H)$ as a function of $H$. The same plot for $A_{2\theta}/(\mu_{0}H)$ obtained from the $ac$-plane rotation is also shown in Fig.\,\ref{tau_H}(a). While a conventional paramagnetic $H$-linear dependence is seen in $A_{2\theta}/(\mu_{0}H)$, $A_{2\phi}/(\mu_{0}H)$ shows a highly unusual magnetic response. We note that the magnitude of $A_{2\phi}/(\mu_{0}H)$ is of the order of a few G, which considerably exceeds the upper limit of remanent difference between the in-plane and $c$ axis magnetization. This is another indication that any possible remanent effects of FM impurities cannot be associated with the observed twofold symmetry in the in-plane $\phi$ scans. We note that this magnitude of $A_{2\phi}/(\mu_{0}H)$ corresponds to the difference of the magnetizations along $[110]$ and $[\bar{1}10]$ directions $M[110]-M[\bar{1}10]$ and hence is a measure of the distribution of in-plane magnetic induction. We infer that this magnitude ($\sim 1$\,G at 4\,T) is consistent with the anomalous broadening of the nuclear-magnetic-resonance (NMR) spectrum observed in the hidden order phase for $\bm{H}\perp c$ \cite{Mat01,Tak07,Ber01}, which is also determined by the field distribution. 

The result of non-trivial field dependence of $A_{2\phi}/(\mu_{0}H)$ indicates that the in-plane magnetizations $M[110](H)$ and $M[\bar{1}10](H)$ have different (non-linear) field dependencies. The non-vanishing $A_{2\phi}/(\mu_{0}H)$ as $\mu_{0}H\to 0$\,T also suggests that the newly found in-plane anisotropy is not induced by magnetic field, but is likely to be an intrinsic property of the (zero-field) ground state of the hidden-order phase. Such a non-vanishing behavior in the zero-field limit may be associated with possible time reversal symmetry breaking. Further measurements sensitive at low fields are important to address this issue.

\section{Constraints on the hidden order parameter}

The broken fourfold rotational symmetry imposes strong constraints on theoretical models which attempt to explain the hidden order \cite{Pep11,Fuj11,Tha11,Opp11,Ike11,Rau12,Cha11}. Based on the observation of the twofold term in the magnetic torque, Thalmeier and Takimoto \cite{Tha11} made the Landau theory analysis for the symmetry calcification of the order parameters allowed in URu$_2$Si$_2$. The tetragonal symmetry group $D_{4h}$ of URu$_2$Si$_2$ has three nontrivial irreducible representations ($A_2$, $B_1$ and $B_2$) and one doubly degenerate $E$ representation. Each representation has the even ($+$) and odd ($-$) parities with respect to time reversal. Among these symmetries, they found that only the $E$-type order parameters can produce the twofold oscillations consistent with the torque experiments. For degenerate $E$ representation there are two possible solutions for the hidden order parameter characterised by the vector $\bm{\eta}=(\eta_a,\eta_b)$. The observed non-zero $\chi_{ab}$ indicates $\bm{\eta}\propto (\pm1,\pm1)$, which leads to two domain types responsible for the phase of the twofold oscillations through the sign of $\eta_a\eta_b$. The symmetry of the hidden order parameter is therefore restricted to $E^{+}$- or $E^{-}$-type, depending on whether the time reversal symmetry is preserved or not. 

Recently Fujimoto proposed the spin nematic order as a candidate for the hidden order \cite{Fuj11}, which breaks the fourfold symmetry in the spin space. This may be categorized in the $E^{+}$ symmetry in the above representation, but involves momentum dependence of the order parameter that may have nodes. He found that this order parameter can explain the temperature dependence of the $\chi_{ab}$ obtained by the torque experiments in a semi-quantitative way. 

More recently, Ikeda {\it et al.} used {\it ab initio} calculations to compare the multipole fluctuations in URu$_2$Si$_2$ \cite{Ike11}. From the complete set of calculations for all the allowed symmetries, they conclude that the $E^{-}$-type rank-5 dotriakontapole order is the most stable order that emerges at low temperatures in URu$_2$Si$_2$. They also found that the antiferromagnetic state is nearly degenerate in energy, which can naturally explain the pressure induced phase transition from the hidden-order phase to the antiferromagnetic state (Fig.\,\ref{TP}) as a spin-flop transition of pseudo-spins from in-plane to out-of-plane direction. The same order has also been proposed by Rau and Kee \cite{Rau12}, who found that the calculated temperature dependence $\chi_{ab}$ in this ordered state is consistent with the experiment. Another $E^{-}$-type order parameter, which has some similarlity to the spin nematic order, has been proposed by Chandra {\it et al.} \cite{Cha11}, who named as a `hastatic' (Latin: {\it spear}) order. Such $E^{-}$ orders break the $C_4$ symmetry as well as the time reversal symmetry. Thus the experimental verifications on the time reversal symmetry breaking are highly desired.

\section{Concluding remarks}

We have reviewed several recent advances on the experimental and theoretical studies of URu$_2$Si$_2$, focusing on the magnetic torque experiments that indicate the fourfold rotational symmetry breaking in the hidden-order phase. Electronic states that break crystal rotational symmetry without showing ordered moments have also been suggested in various strongly correlated electron systems \cite{Kivelson98}, including Sr$_{3}$Ru$_{2}$O$_{7}$ \cite{Bor07}, high-$T_c$ cuprates \cite{Voj09}, and iron pnictides \cite{Kasahara12}, which are discussed in terms of stripe or nematic orders. The finding of a directional electronic state in a heavy-fermion material is another example of such exotic order in correlated matter. The broken $C_4$ symmetry imposes strong constraints on the symmetry of the hidden order parameter to the two-component $E$ representation. The determination of the order parameter requires experimental input as to whether time reversal symmetry is also broken or not. Information on the electronic structure in the hidden order phase is another important factor in the identification of the hidden order parameter. Recent cyclotron resonance experiments allow the full determination of the effective masses on the main Fermi surface sheets \cite{Ton11}. Studies of the Fermi surface structure for the proposed order parameters are a promising way to resolving the long-standing hidden order mystery.

\section*{Acknowledgments}

This review is based on the work in collaboration with Y. Haga, H. Ikeda,  T.\,D. Matsuda, R. Okazaki, Y. Onuki, H.\,J. Shi, and E. Yamamoto. We thank D. Aoki, K. Behnia, P. Chandra, P. Coleman,  S. Fujimoto, R. Ikeda, K. Ishida, H. Harima, Y. Kasahara, G. Knebel, Y. Kohori, Y. Kuramoto, H. Kusunose, T. Sakakibara, H. Shishido, M. Sigrist,  T. Takimoto, P. Thalmeier, A. Schofield, H. Yamagami, and Y. Yanase for helpful discussions. This work was supported by Grant-in-Aid for the Global COE program ``The Next Generation of Physics, Spun from  Universality and Emergence", Grant-in-Aid for Scientific Research on Innovative Areas ``Heavy Electrons'' (No.\,20102002, 20102006, 23102713) from MEXT of Japan, and KAKENHI from JSPS.

\end{document}